\documentclass[cpp,fleqn]{w-art}
\usepackage{times}
\usepackage{w-thm}
\usepackage{amssymb}
\newcommand{\ve}[1]{\mathbf{#1}}

\usepackage[]{graphicx}
\begin{document}
%%    The information for the title page will be placed between
%%    \begin{document} and \maketitle. The order of most entries
%%    is determined by the class file and can not be changed by
%%    rearranging them. The maketitle command follows after the
%%    abstract.
%%
%%    Most of the following commands will be completed by the publisher.
%%
%%    The copyrightyear is defined in the .clo file as the first argument
%%    of the copyrightinfo command. If the copyrightyear differs from that
%%    value it might be adjusted by the following definition:
%%
%% \renewcommand{\copyrightyear}{2003}% uncomment to change the copyrightyear.
%%
\DOIsuffix{theDOIsuffix}
%%
%% issueinfo for header and copyright line
\Volume{}
\Issue{}
\Month{}
\Year{}
%%
%%    First and last pagenumber of the article. If the option
%%    'autolastpage' is set (default) the second argument may be left empty.
\pagespan{1}{}
%%
%%    Dates will be filled in by the publisher. The 'reviseddate' and
%%    'dateposted' (Published online) entry may be left empty.
\Receiveddate{\today}
%\Receiveddate{20 May 2010}
%\Reviseddate{}
%\Accepteddate{}
%\Dateposted{}
%%

%% \pretitle{Editor's Choice}

%% We have a short and a long form for the title. The short form
%% (optional argument) goes into the running head.

\title[Kinetics of Photon Radiation]{Kinetics of Photon Radiation off an
e$^-$e$^+$- Plasma created from the Vacuum in a Strong Laser Field}

%% Please do not enter footnotes or \inst{}-notes into the optional
%% argument of the author command. The optional argument will go into
%% the header.  If there is only one address the marker \inst{x} may be
%% omitted.

%% Information for the first author.
\author[D. B. Blaschke]{D. B. Blaschke\inst{1,2}}
\address[\inst{1}]{Institute for Theoretical Physics,  University of
Wroc{\l}aw, 50-204 Wroc{\l}aw, Poland}
\address[\inst{2}]{Bogoliubov Laboratory for Theoretical Physics,
   Joint Institute for Nuclear Research, 141980 Dubna, Russia}
%%    Information for the second author
\author[G. R\"opke]{G. R\"opke\inst{3}}
\address[\inst{3}]{Institut f\"ur Physik,  Universit\"at Rostock,
D-18051 Rostock, Germany}
\author[S. M. Schmidt]{S. M. Schmidt\inst{4,5}}
\address[\inst{4}]{Forschungszentrum J\"ulich GmbH, D-52428 J\"ulich, Germany}
\address[\inst{5}]{Technische Universit\"at Dortmund, Fakult\"at Physik \&
DELTA, D - 44221 Dortmund, Germany}
%%    Information for the third author
\author[S. A. Smolyansky]{S. A. Smolyansky\footnote{e-mail: {\sf smol@sgu.ru}}
\inst{6}}
\address[\inst{6}]{Saratov State University, 410026 Saratov, Russia}
\author[A. V. Tarakanov]{A. V. Tarakanov%\footnote{Third author footnote.}
\inst{6}}
%%
%%    \dedicatory{This is a dedicatory.}
\begin{abstract}
We consider the one-photon annihilation mechanism in a electron - positron
quasiparticle plasma (EPP) created from the vacuum in a strong subcritical
laser field due to the dynamical Schwinger mechanism.
On the basis of a kinetic theory approach we show that the secondary photons
have a radiation spectrum proportional to $1/k$ (flicker noise).
This effect is very small for EPP excitations in the optical spectrum
but can reach quite observable values in the $\gamma$ - ray region.
\end{abstract}
%% maketitle must follow the abstract.
\maketitle                   % Produces the title.

%% If there is not enough space inside the running head
%% for all authors including the title you may provide
%% the leftmark in one of the following three forms:

%% \renewcommand{\leftmark}
%% {First Author: A Short Title}

%% \renewcommand{\leftmark}
%% {First Author and Second Author: A Short Title}

%% \renewcommand{\leftmark}
%% {First Author et al.: A Short Title}

%% \tableofcontents  % Produces the table of contents.

\section{Introduction}

%\noindent
%{\bf Introduction.}
The effect of vacuum particle creation in a strong electric field was predicted
long ago \cite{SHS} but is not confirmed experimentally up to now.
Theoretical research on this effect in the framework of kinetic theory
\cite{SBR} has shown the existence of a quasiparticle plasma of high density
already in subcritical electric fields $E\ll E_c=m^2/e$  \cite{Vin01}.
Such a quasiparticle plasma does not survive after switching off the external
field in contrast to the real (observed) plasma that remains after applying
the pulse of a strong field $E\geq E_c$ \cite{PNN}, see also
\cite{Alkofer:2001ik} for the case $E\sim E_c$ .
Different proposals have been made for observing the quasiparticle  EPP
generated in the focal spot of  counter propagating laser beams
(see, e.g., Refs.~\cite{BIP,EPJD} and literature cited therein).

One of these effects is based on the possibility of one-photon and two-photon
annihilation of the quasiparticle pairs.
The first reassuring estimations were performed on the basis of the S-matrix
formalism for modern high-power lasers with an intensity of
$I\sim 10^{20}$ W/cm$^2$  \cite{PRL}.
However, there are some doubts in the validity of the S-matrix methods for the
short-lived quasiparticles in the presence of a strong time dependent electric
field.
This stimulates the development of more adequate methods such as a kinetic
description of the EPP, taking into account also its photon component.
The first steps in the direction of including the photon sector of the EPP
into the kinetic description were done in the works \cite{CI,BCA}.

In present work we develop the kinetic theory of the EPP photon sector on the
basis of the one-photon annihilation process which is no longer forbidden
in the presence of strong fields \cite{Rit}.
This requires a derivation of the second level equations of the BBGKY chain.
In the case of an oscillating electric field we show that the photon production
rate is determined by joint action of ponderomotive forces and multi-photon
processes.
The spectrum of the annihilation photons has the $1/k$ behaviour of flicker
noise.
This effect is very small for optical excitations of the EPP but can reach
quite observable values in the $\gamma$ - ray region.
We calculate  the intensity of such processes on the basis of the
"photon count method" \cite{BIP}.
We have the methodical basis for more advanced calculations of the photon
production rate for the two-photon annihilation mechanism requiring the fourth
level equations of the BBGKY chain.

Let us stress that the present estimates are actual especially in view of the
planned experiment for testing the subcritical EPP production in vacuum
with the Astra-Gemini high power laser \cite{GG}.

%\subsection{First subsection}

\section{One-photon annihilation process}
%\noindent
%{\bf One-photon annihilation process.}
An external electric field generates a vacuum instability due to EPP creation
accompanied by the appearance of internal currents and electromagnetic fields
(backreaction  problem).
Quantum fluctuations of this internal field are interpreted as photon
excitations, which can leave the active zone (focal spot) and can be
detected.
Below the KE of the photon component will be examined on the basis of the
one-photon annihilation mechanism, thus going beyond the investigation of
the KE in the fermionic quasiparticle sector in previous works
\cite{Kiel04,Fil08,PS}.

The single-time, two-point photon correlation function in momentum space is
\begin{equation}\label{25e}
    F_{rr'}(\ve{k},\ve{k}',t)=
\langle A^{(+)}_{r}(\ve{k},t)A^{(-)}_{r'}(\ve{k}',t)\rangle~,
\end{equation}
with the photon vacuum $\langle A_r ^{(\pm)}(\ve{k},t) \rangle=0$.
Let us write the first equation of the BBGKY hierarchy
{\small
\begin{eqnarray}
\label{27e}
   \dot{F}_{rr'}(\ve{k},\ve{k}',t) &=&
ie\int \frac{d^3 p_1 d^3 p_2}{\sqrt{(2\pi)^3}}\Bigl\{\frac{
\delta(\ve{p}_1 -\ve{p}_2 -\ve{k})}{\sqrt{2k}}
 [\bar{u}v]^{r}_{\beta\alpha}(\ve{p}_1 ,\ve{p}_2 ,\ve{k};t)
\langle b^{+}_{\beta}(-\ve{p}_{2},t)a^+ _{\alpha}(\ve{p}_1 ,t)
A_{r'}^{(-)} (\ve{k}',t)\rangle \nonumber \\
& &+ \frac{\delta(\ve{p}_1 -\ve{p}_2 +\ve{k}')}{\sqrt{2k'}}
[\bar{v}u]^{r'}_{\beta\alpha}(\ve{p}_1 ,\ve{p}_2 ,\ve{k}';t)
\langle b_{\alpha}(-\ve{p}_{1},t)a_{\beta}(\ve{p}_2 ,t)A_{r} ^{(+)} (\ve{k},t)
\rangle \Bigr\}~,
\end{eqnarray}
}
where $a_\alpha, b_\beta, A_{r}^{(-)}$
($a^+_\alpha, b^+_\beta, A_{r}^{(+)}$) denote the annihilation (creation)
operators for electrons, positrons and photons, respectively \cite{BCA}.
For obtaining a closed photon KE a truncation procedure for the correlators
entering this equation is required.
The equations of the second order for the correlators in Eq.~(\ref{27e}) can
be obtained using the Heisenberg-like equations of motion, e.g.,
{\small
\begin{eqnarray}
\label{29e}
\biggl\{\frac{\partial}{\partial t}&+&
i[\omega(\ve{p}_1,t)+\omega(\ve{p}_2,t)-k]\biggr\}
\langle b_{\alpha}(-\ve{p}_{1},t)a_{\beta}(\ve{p}_2 ,t)A_{r} ^{(+)} (\ve{k},t)
\rangle \nonumber \\
&=&-ie\int\frac{d^3p'}{\sqrt{(2\pi)^3}}\frac{d^3k'}{\sqrt{2k'}}
\Bigl\{ \delta(\ve{p}' -\ve{p}_1 +\ve{k}')
\nonumber\\
&\times& \left[[\bar{u}v]^{r'}_{\alpha\beta'}(\ve{p}' ,\ve{p}_1 ,\ve{k}';t)
\langle a^{+}_{\beta'}(\ve{p}',t)a_{\beta}(\ve{p}_2 ,t)A_{r'}
(\ve{k}',t)A_{r} ^{(+)} (\ve{k},t)\rangle \right.
\nonumber \\
 &+&\left. [\bar{v}v]^{r'}_{\alpha\beta'}(\ve{p}' ,\ve{p}_1 ,\ve{k}';t)
\langle b_{\beta'}(-\ve{p}',t)a_{\beta}(\ve{p}_2 ,t)A_{r'}
(\ve{k}',t)A_{r} ^{(+)} (\ve{k},t)\rangle \right]
\nonumber \\
 &-&\delta(\ve{p}_2 -\ve{p}' +\ve{k}')
 \cdot \left[[\bar{u}u]^{r'}_{\beta'\beta}(\ve{p}_2 ,\ve{p}' ,\ve{k}';t)
\langle b_{\alpha}(-\ve{p}_1,t)a_{\beta'}(\ve{p}' ,t)A_{r'}
(\ve{k}',t)A_{r} ^{(+)} (\ve{k},t)\rangle \right.
\nonumber \\
 &+& \left.[\bar{u}v]^{r'}_{\beta'\beta}(\ve{p}_2 ,\ve{p}' ,\ve{k}';t)
\langle b_{\alpha}(-\ve{p}_1,t)b_{\beta'}^+ (-\ve{p}' ,t)A_{r'}
(\ve{k}',t)A_{r} ^{(+)} (\ve{k},t)\rangle \right]\Bigr\}
\nonumber \\
 &+& S^r_{\alpha\beta}(\ve{p}_1 ,\ve{p}_2 ,\ve{k};t)
+ U^r _{\alpha\beta}(\ve{p}_1 ,\ve{p}_2 ,\ve{k};t)~.
\end{eqnarray}
}
On the r.h.s. of this equation there is a set of terms defined by the vacuum
polarization effects in the presence of the quantized electromagnetic field
(summed up in $U^{(r)}$).
The other set of terms (summed up in $S^{(r)}$)
is connected with different transformations in the $e^- e^+\gamma$- plasma
without participation of the quasiparticle photon excitations.
These processes have an indirect influence on the photon distribution
due to the secondary momentum redistribution of quasiparticles by means of
collisions.
These $U^r$ and $S^r$ terms will be omitted below (the approximation A1).

The remaining processes are written out explicitly in Eq.~(\ref{29e}).
The first and the fourth terms describe the mutual influence of the electron
and positron quasiparticle excitations and photons.
Only these last terms will be kept subsequently in the first step
(the approximation A2).
Some basis for such selection is the following reasoning:
in RPA the truncation of correlators of the type
\begin{equation}
\label{32e}
\langle a^{+}_{\beta'}(\ve{p}',t)a_{\beta}(\ve{p}_2 ,t)A_{r'}
(\ve{k}',t)A_{r} ^{(+)} (\ve{k},t)\rangle \simeq
\langle a^{+}_{\beta'}(\ve{p}',t)a_{\beta}(\ve{p}_2 ,t)\rangle
\langle A_{r'}  (\ve{k}',t)A_{r} ^{(+)} (\ve{k},t)\rangle
\end{equation}
leads to a nonvanishing contribution already in the lowest order of
perturbation theory.
In this approximation Eq.~(\ref{29e}) gets the form
{\small
\begin{eqnarray}
\label{33e}
\biggl\{\frac{\partial}{\partial t}&+&
i[\omega(\ve{p}_1,t)+\omega(\ve{p}_2,t)-k]\biggr\}
\langle b_{\alpha}(-\ve{p}_{1},t)a_{\beta}(\ve{p}_2 ,t)A_{r} ^{(+)}
(\ve{k},t)\rangle~ \simeq \nonumber\\
&-&ie\int \frac{d^3 p'}{\sqrt{(2\pi)^3}} \frac{d^3 k'}{\sqrt{2k'}}
\Bigl\{ \delta(\ve{p}' -\ve{p}_1 +\ve{k}')
 [\bar{u}v]^{r'}_{\alpha\beta'}(\ve{p}' ,\ve{p}_1 ,\ve{k}';t) \nonumber  \\
 &\times& \langle a^{+}_{\beta'}(\ve{p}',t)a_{\beta}(\ve{p}_2 ,t)A_{r'}^{(-)}
(\ve{k}',t)A_{r} ^{(+)} (\ve{k},t)\rangle
~-~ \delta(\ve{p}_2 -\ve{p}' +\ve{k}') \nonumber \\
&\cdot& [\bar{u}v]^{r'}_{\beta'\beta}(\ve{p}_2 ,\ve{p}' ,\ve{k}';t)
\langle b_{\alpha}(-\ve{p}_1,t)b_{\beta'}^+ (-\ve{p}' ,t)A_{r'}^{(-)}
(\ve{k}',t)A_{r} ^{(+)} (\ve{k},t) \rangle \Bigr\}~.
\end{eqnarray}
}
The processes of the instantaneous radiation of two photons was omitted here,
i.e., in Eq.~(\ref{29e}) the substitution
$A_{r'}(\ve{k}',t)\rightarrow A_{r'}^{(-)}(\ve{k}',t)$ has been made
(the approximation A3).
Their inclusion requires the consideration of the next (the third) level
equation of the BBGKY chain under the correlators of the type
$$
\langle a_{\beta'}^+ (\ve{p}',t)a_{\beta}(\ve{p}_2 ,t)A_{r'}^{(+)}
(\ve{k}',t)A_{r} ^{(+)} (\ve{k},t)\rangle~.
$$
Let us remark that the two-photon annihilation process with the correlator of
the type (the second and the third terms on the r.h.s. of Eq.~(\ref{29e}))~
$
\langle b_{\beta'}(-\ve{p}',t)a_{\beta}(\ve{p}_2 ,t)A_{r'}^{(+)}
(\ve{k}',t)A_{r}^{(+)} (\ve{k},t)\rangle~
$
is also found among the processes omitted in this section.
Thus, the selected simplest approximation here corresponds to taking into
account the one-photon annihilation process only.
Let us rewrite Eq.~(\ref{33e}) in integral form
{\small
\begin{eqnarray}
\label{40e}
\langle b_{\alpha}(-\ve{p}_{1},t)a_{\beta}(\ve{p}_2 ,t)A_{r}^{(+)}
(\ve{k},t)\rangle
&=& -\frac{ie}{\sqrt{2k}} \int_{t_0}^t \frac{dt'}{\sqrt{(2\pi)^3}}
\exp\left\{ -i\int_{t'}^{t}d\tau[\omega(\ve{p}_1,\tau)
+\omega(\ve{p}_2,\tau)-k]\right\}
\nonumber\\
&\times& \left\{[\bar{u}v]^{r}_{\alpha\beta'}(\ve{p}_2 ,\ve{p}_1 ,\ve{k};t')
 f_{\beta\beta'}(\ve{p}_2,t')
  - [\bar{u}v]^{r}_{\beta'\beta}(\ve{p}_2 ,\ve{p}_1 ,\ve{k};t')
f_{\beta'\alpha}^c(\ve{p}_1,t')\right\}
\nonumber \\
&\times&
\{1+F_r (\ve{k},t')\}\delta(\ve{p}_2 -\ve{p}_1 +\ve{k})~.
\end{eqnarray}
}
In order to obtain Eq.~(\ref{40e}), the following truncation procedure
(the approximation A4) has been applied
$$
\langle a^{+}_{\beta'}(\ve{p}',t')a_{\beta}(\ve{p}_2 ,t')A_{r'}^{(-)}
(\ve{k}',t')A_{r} ^{(+)} (\ve{k},t') \rangle
=  \delta_{rr'}\delta(\ve{p}'-\ve{p}_2)\delta(\ve{k}-\ve{k}')
f_{\beta\beta'}(\ve{p}_2,t')\{1+F_r (\ve{k},t')\}~,
$$
and use was made of the relation~
$
\langle a^{+}_{\beta}(\ve{p},t)a_{\alpha}(\ve{p}' ,t)\rangle
=\delta(\ve{p}-\ve{p}')f_{\alpha\beta'}(\ve{p},t)~.
$
The relation (\ref{40e}) combined with Eq.~(\ref{27e}) and the system of
equations of the fermion sector leads to a formally closed KE system.
However, keeping in mind the estimation of the radiated photon spectrum,
we will introduce a set of the additional simplifying assumptions.
An essential simplification is reached if we neglect the spin effect in
Eq.~(\ref{40e}), i.e.,
$f_{\alpha\beta}\rightarrow f\delta_{\alpha\beta}$ (the approximation A5).
This leads to the following KE of non-Markovian type
\begin{eqnarray}
\label{43e}
\dot{F}(\ve{k},t) &=& \frac{e^2}{2(2\pi)^{3}k}\int d^3 p  \int_{t_0}^{t}dt'
   \cos\left\{ \int_{t'}^{t}d\tau
[\omega(\ve{p},\tau)+\omega(\ve{p}+\ve{k},\tau)-k]\right\}
\nonumber \\
  & &\times K(\ve{p},\ve{p} +\ve{k},\ve{k}|t,t')
  \{f(\ve{p} +\ve{k},t')+f(\ve{p},t')-1\}[1+F_r (\ve{k},t')]~,
\end{eqnarray}
where the kernel $K$ is defined in the nonstationary spinor basis
\cite{Kiel04,Fil08,PS}  as the spinor construction
\begin{equation}
\label{44e}
 K(\ve{p}_1 ,\ve{p}_2,\ve{k}|t,t')
  =[\bar{v}u]^{r}_{\beta\alpha}(\ve{p}_1 ,\ve{p}_2 ,\ve{k};t)
   [\bar{u}v]^{r}_{\alpha\beta}(\ve{p}_2 ,\ve{p}_1 ,\ve{k};t')~.
\end{equation}
We assume that the photon distribution has not a fixed direction of
polarization, i.e.,
  $
    F_1=F_2=F
  $
(the approximation A6).
It has also been taken into account that $f^c =1-f$ due to electroneutrality
of the vacuum.

\section{The case of a weak external field}
%\noindent
%{\bf The case of a weak external field.}
If $E(t)\ll E_c$, it can be expected that the density of the observable
photons is small, $F(\ve{k},t)\ll 1$, and thus their distribution function
can be neglected on the r.h.s. of Eq.~(\ref{43e}), the photon production rate.
%\begin{eqnarray}
%\dot{F}(\ve{k},t)
%&=& \frac{e^2}{2(2\pi)^{3}k}\int d^3 p  \int_{t_0}^{t}dt'
%    K(\ve{p},\ve{p} +\ve{k},\ve{k}|t,t')
%    \{f(\ve{p},t')+f(\ve{p} +\ve{k},t')-1\}
%\nonumber \\
%& & \cos\left\{ \int_{t'}^{t}d\tau
%    [\omega(\ve{p},\tau)+\omega(\ve{p}+\ve{k},\tau)-k]\right\}.
%\label{44_1e}
%\end{eqnarray}
Thus, the photon generation is defined, in the first place, by  the statistical
factor
$
    1-f(\ve{p},t')-f(\ve{p} +\ve{k},t')~,
$
which describes the quasiparticle EPP as the material medium generating
real photons and the fast oscillating factor with the phase
\begin{equation}
\label{44_3e}
\Phi(t,t')=\int_{t'}^td\tau[\omega(\ve{p},\tau)+\omega(\ve{p}+\ve{k},\tau)-k]~,
\end{equation}
which is defined by the dispersion relation of the partners involved in the
one-photon annihilation process.

The kernel (\ref{44e}) is a slowly varying function of the momentum arguments
$\ve{p}_1,\ve{p}_2$ and some weak oscillations in $t$ and $t'$.
For a gross estimate one can neglect the retardation effect by setting $t=t'$.
Then one can use the well-known rules of summation over the spin indices to
obtain
$
       K(\ve{p},\ve{p}-\ve{k};t,t')\rightarrow K_0 \sim 1.
  $
Apparently, this leads to a divergence of the momentum space integral on the
r.h.s. of Eq.~(\ref{43e})
%(\ref{44_1e})
from the ``1'' in the statistical factor.
Subtracting  the corresponding contribution we obtain the renormalized
production rate of real photons
\begin{equation}
  \dot{F}(\ve{k},t) =\frac{e^2 K_0}{2(2\pi)^{3}k}\int d^3 p  \int_{t_0}^{t}dt'
   \{f(\ve{p},t')+f(\ve{p} +\ve{k},t')\}\cos\Phi(t,t')~.
\label{46e}
\end{equation}

The frequency dependence $F(\ve{k},t)\sim 1/k$ in the infrared region
$k\ll m$  (see Eq.~(\ref{46e})) is the characteristic behaviour of flicker
noise (e.g., \cite{Kus}), which arises in the present case from the
fluctuations of vacuum oscillations \cite{SarRev}.
Technically, this result is a direct consequence of the decomposition for the
electromagnetic field operators in the region of small $k$.
The phase (\ref{44_3e}) is always very large since the mismatch
\begin{equation}\label{50e}
        \omega(\ve{p},\tau)+\omega(\ve{p}+\ve{k},\tau)-k\sim 2m
\end{equation}
is very large and the energy conservation law is violated for the one-photon
annihilation process in the absence of an external field.
That is the reason why the radiation of a real photon can be interpreted as
the multiphoton process.
The photon number $N_{\nu}$ from the photon condensate of the external
quasiclassical photons with the frequency $\nu$ can be estimated if it has to
compensate the mismatch (\ref{50e}) in the energy of $N_{\nu}$ quasiclassical
photons.
This leads to $N_{\nu}\sim 2m/\nu$.
For optical lasers it is a huge number and therefore such a fluctuation event
is very rare.
This conclusion about the role of multiphoton processes is correlated with the
analysis of the absorption coefficient of the EPP created from the vacuum in
the infrared region \cite{BIP}.

In order to generalize this result to the case of arbitrary energies for the
real photons, we perform the Fourier decomposition of the functions on the
r.h.s. of Eq.~(\ref{46e}) (method of photon count \cite{BIP}).
This is a rather complicated problem because the time scales are defined by
two parameters with dimension of energy (mass $m$ and the frequency $\nu$ of
the external periodical field) under the arbitrary third parameter $k$, the
energy of the real photon.
The basic idea is the compensation of the mismatch (\ref{50e}) with help of
the multiphoton process in order to eliminate  the beating of the phase
(\ref{44_3e}) (the resonance approximation).
In Eq.~(\ref{46e}) the source of the photon excitations with the energy $k$
is the fermion distribution function which is an implicit function defined by
the KE in the fermion sector.

The other  photon source is the fastly oscillating factor with the phase
(\ref{44_3e}) which depends on the parameters of the external field.
For subcritical fields $E\ll E_c$ we can write approximately
($\omega_0(\ve{p})=\omega(\ve{p},t)|_{A=0}$)
$
   \omega(\ve{p},t)\simeq \omega_0(\ve{p})
	-\frac{e\ve{E}_0 \ve{p}}{\nu\, \omega_0(\ve{p})}\cos\nu t ,\,
   \frac{e|\ve{E}_0 \ve{p}|}{\nu\, \omega_0^2(\ve{p})}\ll 1~,
$
if $\ve{A}(t)=(\ve{E}_0/\nu ) \cos\nu t~.$
Then the phase (\ref{44_3e}) is given by
\begin{equation}
\label{52e}
        \Phi(t,t')\simeq \Omega_0(\ve{p},\ve{k})(t-t')
		+a(\ve{p},\ve{k})[\sin\nu t-\sin\nu t'],
\end{equation}
where the mismatch is
\begin{equation}\label{53e}
        \Omega_0(\ve{p},\ve{k})=\omega_0(\ve{p})+\omega_0(\ve{p}+\ve{k})-k
\end{equation}
and
$
a(\ve{p},\ve{k})=-{e}/{\nu^2}\left\{[\ve{E}_0\ve{p}]/{\omega_0(\ve{p})}
+ [\ve{E}_0(\ve{p}+\ve{k})]/{\omega_0(\ve{p}+\ve{k})}\right\}.
$
The fastly oscillating function in Eq.~(\ref{46e}) allows then the following
representation
\begin{equation}
\label{55e}
    \cos\Phi(t,t')\simeq e^{i\Omega_0(t-t')}
	\sum_{n,n'}J_n(a)J_{n'}(a)e^{i\nu(nt-n't')}+c.c.~,
\end{equation}
where $J_n (a)$ is the Bessel function of order $n$.
Thus, the representation (\ref{55e}) contains two high frequency harmonics
$\omega_0(\ve{p})$ and $\omega_0(\ve{p}+\ve{k})$ (see Eq.(\ref{53e})), the set
of the low frequency harmonics $n\nu$ and the $k$- harmonic corresponding to
the radiating photon.
This structure of the Eq.~(\ref{55e}) and the discussion of the mismatch
(\ref{50e}) suggest the following twofold decomposition of the fermion
distribution function ($f_{-n-l}=f_{nl}^*$):
\begin{equation}
\label{57e}
    f(\ve{p},t)=\sum_{n,l}f_{n,l}(\ve{p})e^{in\nu t+imlt}~,
\end{equation}
which has two modes: the soft ("breathing") one with the frequencies $n\nu$ and
the hard ("trembling") one with the harmonics $lm$.
Substituting Eqs.~(\ref{55e}) and (\ref{57e}) into Eq.~(\ref{46e}) leads to the
relation
{\small
\begin{equation}
   \dot{F}(\ve{k},t) =\frac{i e^2 K_0}{4(2\pi)^{3}k}
   \sum_{n_1,l}\sum_{n,n'}\int d^3 p  J_n(a)J_{n'}(a)
   \{f_{n_1 l}(\ve{p})+f_{n_1 l}(\ve{p} +\ve{k})\}
   \frac{\exp\left\{ i[lm+\nu(n+n_1-n')]t\right\}}
   {\Omega_0-lm-\nu(n_1-n')+i\varepsilon}~.
\label{58e}
\end{equation}
}
The shift $i\varepsilon$ in the complex plane is a consequence of the adiabatic
hypothesis garanteeing the convergence of the integral for
$t_0\rightarrow-\infty$.

Let us use now the resonance approximation which allows to select from the sum
on r.h.s. of Eq.~(\ref{58e}) the constant component
$
    lm+(n+n_1 -n')\nu=0.
$
At $l=0$ the decomposition (\ref{58e}) contains the basic low frequency
breathing modes only.
Below we will consider this case only (the case $l=1$ must be investigated
separately also).
The photon production rate (\ref{58e}) has then the form
{\small
\begin{eqnarray}
  \dot{F}(\ve{k})& =&\frac{ie^2 K_0}{4(2\pi)^{3}k }\sum_{n,n'}\int d^3 p
	J_n(a)J_{n'}(a)
       \frac{f_{n'-n,0}(\ve{p}) +f_{n'-n,0}(\ve{p}+\ve{k})}
	{\Omega_0+n\nu+i\varepsilon}
\nonumber\\
    &=&\frac{\pi e^2 K_0}{4(2\pi)^{3}k }\sum_{n,n'}\int d^3 p
	J_n(a)J_{n'}(a)
      \left\{f_{n'-n,0}(\ve{p}) +f_{n'-n,0}(\ve{p}+\ve{k})\right\}
	\delta(\Omega_0+n\nu)~.
\label{60e}
\end{eqnarray}
}
Using the textbook formula
{\small
$
    \delta[\phi(x)]=\sum_i\left\{|\phi'(x_i)|\right\}^{-1}\delta(x-x_i),\,
\phi(x_i)=0,
$
}
we obtain
\begin{equation}
  \dot{F}({k}) =\frac{\alpha K_0}{2 k }\sum_{n,n'}\xi(k,n)
	J_n(a_0)J_{n'}(a_0)
      \left\{f_{n'-n,0}({p}_0) +f_{n'-n,0}({p}_0+{k})\right\}~,
\label{63e}
\end{equation}
where $a_0=a(p_0)$, $\alpha=e^2/4\pi$ and according to Eq.~(\ref{53e})
\begin{equation}
\label{64e}
     \xi(k,n)=p_0^2/|\Omega'_0(p_0)|
	=p_0\frac{\omega_0(p_0)\sqrt{\omega^2_0(p_0)+k^2}}
	{\omega_0(p_0)+\sqrt{\omega^2_0(p_0)+k^2}}
\end{equation}
and $p_0$ is the positive root of the equation $\Omega_0(p_0)-n\nu=0$~,
\begin{equation}\label{65e}
    p_0=\left\{\frac{(n\nu)^2(n\nu-2k)^2}{4(n\nu -k)^2}-m^2\right\}^{1/2}\geq 0
\end{equation}
and
\begin{equation}\label{66e}
   n= \frac{1}{\nu}\left\{\omega_0(\ve{p})+\omega_0(\ve{p}+\ve{k})-k \right\}~.
\end{equation}
One can see that $n\in [\omega_0({p}),2\omega_0({p})]$, i.e. $n$ is very large
in the optical region and can be $\sim 1$ for $\nu\sim m$.

Now we take into account the well known property of the fermion distribution
function: the basic breathing mode corresponds to the second harmonic
($n=\pm 2$) in Eq.~(\ref{57e}).
Keeping only these modes ($n'-n=\pm 2$) in Eq.~(\ref{63e}), we obtain
\begin{equation}
  \dot{F}({k}) =\frac{\alpha K_0}{2 k }\sum_{n\geq n_0}\xi(k,n)
	J_n(a_0)[J_{n+2}(a_0)+J_{n-2}(a_0)]
      \left\{f_{2,0}(\ve{p}_0) +f_{2,0}(\ve{p}_0+\ve{k})\right\}~,
\label{66_2e}
\end{equation}
where according to Eq.~(\ref{66e})
\begin{equation}
\label{66_3e}
   n_0=n(p=0)= \frac{1}{\nu}\left\{m+\omega({k})-k \right\}~.
\end{equation}
Leaving in Eq.~(\ref{66_2e}) only the leading term ($n=n_0+1$), we obtain
\begin{equation}
  \dot{F}({k}) =\frac{\alpha K_0}{2 k }\xi(k,n_0+1)
	J_{n_0+1}(a_0)[J_{n_0+3}(a_0)+J_{n-1}(a_0)]
      \left\{f_{2,0}(\ve{p}_0) +f_{2,0}(\ve{p}_0+\ve{k})\right\}~,
\label{66_4e}
\end{equation}
where
\begin{equation}
  \xi(k,n_0+1)={p}_0(n_0+1)\frac{m\omega_0({k})}{m+\omega_0({k})}
\label{66_5e}
\end{equation}
and $p_0(n_0+1)$ can be found from Eqs.~(\ref{65e}) and (\ref{66_3e}).
Using the decompositions of the Bessel functions
%\begin{equation}
$
 J_0(a_0)=1, \quad J_{n_0}(a_0)=(a_0/2)^{n_0}/{n_0}\sim\alpha^{n_0/2}~,
$
%\label{66_6e}
%\end{equation}
valid for $a_0\ll 1$, we obtain the final estimates.
The photon production rate (\ref{66_4e}) in the case of the one-photon
annihilation mechanism is very small $\dot{F}\sim\alpha^{n_0/2}$ for the
optical laser radiation with $n_0\gg 1$, see Eq.~(\ref{66e}).
This means that the excitation of a single observable photon requires a huge
number of optical "laser" photons so that the probability of such process is
negligible.
However, in the $\gamma$- ray region $\nu\sim m$ and then $n_0\sim 1$.
In this case the photon production rate can be quite observable.
For example, for $n_0=1$ (it corresponds to $\nu=2m$) and for soft
photon radiation $k\ll m$ we obtain $\dot{F}\sim \alpha^2$.

%--------------------------------------
\section{Summary \label{sect:summmary}}
%\noindent
%{\bf Summary.}
We have shown that the one-photon annihilation mechanism acting
in the quasiparticle EPP can lead to the excitation of observable photons by
multi-photon processes.
In the infrared region $k\ll m$ the radiation spectrum behaves as $1/k$
(the flicker noise).
The increase $k$ results in a more complicated spectrum, Eq.~(\ref{66_4e}).
In the case of subcritical laser fields $E\ll E_c$ and optical excitations
of the EPP this effect is very small.
However, in the $\gamma$-ray region the photon production rate can increase up
to quite observable values.
In this connection one can expect that the absence of the multi-photon
excitations mechanism in the two-photon annihilation channel can lead to the
generation of an experimentally significant photon production intensity in the
optical laser domain.

\subsection*{Acknowledgement}
%\noindent
%{\bf Acknowledgement.}
D.B.B. and S.A.S. acknowledge financial support of the Forschungszentrum
J\"ulich, and the hospitality at its IKP where this research has been started.

\end{document}